\documentstyle[a4]{article}
\def\l{\langle}
\def\r{\rangle}

\def\dn{\delta n}
\def\dns{\dn^2}

\def\hom{\hbar \omega}

\def\bra{\langle}
\def\ket{\rangle}

\def\vpa{\varphi_a}
\def\vpb{\varphi_b}

\def\hf{ {1 \over 2} }
\def\ban{\delta \phi_{BA}}
\def\me{\delta n_{err}^2}

\begin{document}
\baselineskip=4mm
\begin{flushleft}
\Large\bf
Quantum Measurement and Fluctuations in Nanostructures
\footnote{Published in
{\bf Nanostructures and Quantum Effects}, eds. H. Sakaki and H. Noge
(Springer Series in Material Sciences, Vol.31, 1994)
}
\bigskip

\large\it
A. Shimizu
\bigskip

\normalsize\rm
Institute of Physics, University of Tokyo, Komaba, Tokyo 153, Japan

\end{flushleft}
\bigskip

\noindent{\bf Abstract.}
 Measurement and fluctuations are closely related to each other
in quantum mechanics.
This fact is explicitly demonstrated in the case of
a quantum non-demolition photodetector which is
composed of a double quantum-wire electron interferometer.
\bigskip\medskip

\noindent{\bf 1. Introduction}
\bigskip

\noindent
Recent rapid progress of studies on nanostructures
is opening up possibilities of new measuring  apparatus using  nanostructures.
For example, a tiny change of the electric charge in a nano-scale
region can be detected through a single-electron-tunneling
transistor \cite{SET}.
Another example is a quantum-wire electron interferometer
that works as a quantum non-demolition (QND) photodetector, which
measures the photon number without absorbing photons \cite{Spra}.
The functions of these nanostructure devices are hardly accessible by
conventional devices, thus
make nanostructure devices very attractive.

On the other hand, these devices
stimulate studies on
a very basic problem of physics---
what happens when you measure a quantum system?
To discuss this problem the nanostructure devices are useful
because they allow microscopic analysis of the measuring devices.
As a result,
we can clarify close relationships
among the measurement error, backactions, and fluctuations.
I here demonstrate these things by reviewing
our studies on the quantum-wire QND photodetector.
\bigskip\medskip

\noindent{\bf 2. Quantum-wire QND photodetector}
\bigskip

\noindent
A schematic diagram of the
quantum non-demolition (QND) photodetector \cite{Spra}
is shown in Fig.1. Before going to the full analysis in the following
sections, I here give an intuitive, semi-classical  description \cite{SFOY}
of the operation principle.

The device is
composed of two quantum wires, N and W.
The lowest subband energies (of the $z$-direction confinement)
$\epsilon_a^N$ and $\epsilon_a^W$
of the wires are the same, but the second levels
$\epsilon_b^N$ and $\epsilon_b^W$ are different.
Electrons occupy the lowest levels only.
A $z$-polarized light beam
hits the dotted region. The photon energy $\hbar \omega$ is assumed to
satisfy
$\epsilon_b^W - \epsilon_a^W < \hbar \omega < \epsilon_b^N - \epsilon_a^N$,
so that real excitation does not occur and no photons are absorbed.
However,
the electrons are excited ``virtually" \cite{virtual}, and
the electron wavefunction undergoes a phase shift
between its amplitudes in the two wires.
Since the magnitude of the virtual excitation is proportional to
the light intensity \cite{virtual}, so is the phase shift.
This phase shift modulates the interference currents, $J_+$ and $J_-$.
By measuring $J_\pm$, we can know the magnitude of the phase shift,
from which we can know the light intensity.
Since the light intensity is proportional to the photon number $n$,
we can get information on $n$.
We thus get to know $n$
without photon absorption, i.e.,
without changing $n$; hence the name QND \cite{qndquadra}.
 (More accurate definition of QND will be given in section 7.)
In contrast, conventional photodetectors drastically alter the photon number by
absorbing photons.
Keeping this semi-classical argument in mind,
let us proceed to a fully-quantum analysis.

\begin{minipage}{10cm}
\vspace{100mm}
\noindent
{\bf Fig.~1} A quantum non-demolition
photodetector composed of a double-quantum wire
electron interferometer.
(Taken from \cite{Spra})
\end{minipage}

\bigskip\medskip

\noindent{\bf 3. Quantized light field for a waveguide mode}
\bigskip

\noindent
We assume that the measured light
of frequency $\omega$,
plane polarized in the z direction,
is confined in the x and z directions in a waveguide,
propagating in the y direction with the propagation constant $\beta_\omega$.
A normalized mode function ${\bf u}({\bf r})$ then takes the form
\begin{equation}
{\bf u}({\bf r}) = (0, 0, u({\bf r}) ), \quad
u({\bf r})
= v_\omega(x,z) \exp(i \beta_\omega y)/\sqrt{L_y}.
\label{mode} \end{equation}
Here, $L_y$ is a normalization length, and
$v_\omega(x,z)$ is the lateral mode function.
$u({\bf r})$ is normalized as \cite{GL}
\begin{equation}
\int \epsilon |u|^2 d^3 {\bf r} = 1,
\end{equation}
where $\epsilon$ is the dielectric constant. (The permeability is
unity at the optical frequency.)
The quantized optical electric field in this mode is expressed as \cite{G}
\begin{equation}
\hat{\cal E}({\bf r},t) =
\sqrt{{2 \pi \hom}}
\left[ \hat a  {\bf u}({\bf r}) e^{-i \omega t} + h.c. \right].
\label{E} \end{equation}
The annihilation operator $\hat a$ thus defined is
the one for a freely propagating waveguide mode.
When mirrors are placed at $y = \pm L_y/2$, on the other hand,
the measured light is
confined in all directions, and ${\bf u}({\bf r})$ is then given by
a superposition of Eq.\ (\ref{mode}) with $\pm \beta_\omega$.
Using such ${\bf u}({\bf r})$ in Eq.\ (\ref{E}), we obtain
$\hat a$ for the confined mode \cite{unitary},
and $\hat n \equiv \hat a^\dagger \hat a$ then defines
the photon number {\it in the confined mode} \cite{G}.
We can also treat the case where
the measured light takes a wavepacket form, for which
the ``mode function''
is given by a superposition of ${\bf u}({\bf r}) e^{-i \omega t}$ of Eq.\ (\ref{mode})
over a narrow range of $\omega$.
Replacing ${\bf u}({\bf r}) e^{-i \omega t}$ with this mode function in Eq.\ (\ref{E}),
we obtain $\hat a$ for the wavepacket mode \cite{unitary},
and $\hat n \equiv \hat a^\dagger \hat a$ then defines
the photon number in the wavepacket \cite{G}.
In any case, the number state is  defined by $\hat n | n \r = n | n \r$,
with $n=0, 1, 2, \ldots$, and
any state vector of light in the mode of interest can be expressed as
\begin{equation}
|\psi_{ph} \r = \sum_n a_n | n \r,
\label {phinit} \end{equation}
which we assume for the state before the measurement.

Either of the above three cases can be treated in a similar manner
in the following discussions.
However, since equations become slightly complicated in the wavepacket case,
we hereafter assume the former two cases.
\bigskip\medskip

\noindent{\bf 4. Single quantum-wire structure}
\bigskip

\noindent
Before going to the full analysis, let us consider the simplified case where
the light field interacts with
a  {\it single} electron which is confined in a {\it single} quantum-wire structure.

Assuming for simplicity that the confinement potential in the y direction is
high enough, we can decompose the y dependence of the electron wavefunction;
$\psi_{el}({\bf r},t) = \psi_{el}(x,z,t) Y(y)$.
Hence, we hereafter drop the $y$-subband eigenfunction
$Y(y)$ from equations.

The electron is emitted from the source region, which is in
the thermal equilibrium (of zero temperature, for simplicity).
Hence, no quantum coherence exists between the electron and photons
before they interact.
To describe this fact,
it is convenient to consider that the initial ($t=0$)
wavefunction of the electron takes a wavepacket form;
\begin{equation}
\psi_{el}(x,z,t=0) =  e^{i k x} G(x) \varphi_a(z).
\end{equation}
Here, $\varphi_a$ is the eigenfunction of the lowest level
(which the electron is assumed to occupy) of the $z$ subbands,
and $G$ is a localized function.
When $G$ does not change appreciably on the scale of the Fermi wavelength,
the detailed form of $G(x)$ is irrelevant to the following results.
We will therefore use the simplified notation like
$|\psi_{el} \r =  |\varphi_a \r$.
Combining this with Eq.\ (\ref{phinit}), we write for
the initial state vector of the coupled photon-electron system as
\begin{equation}
|\Psi \r = |\psi_{ph} \r |\psi_{el}\r = \sum_n a_n  |n \r |\varphi_a \r.
\label{init} \end{equation}
Our task is now to investigate its time evolution ---
I will present here only the {\it final} state,
i.e., the state after the photon-electron collision.

Let us work in the Schr\"odinger picture,
in which the optical electric-field operator $\hat{\cal E}({\bf r})$ is given by
$\hat{\cal E}({\bf r},t=0)$ of Eq.\ (\ref{E}).
The Hamiltonian of the {\it coupled photon-electron system}
is given by
\begin{equation}
H = H_{ph} + H_{el} + H_I, \quad H_I = - e {\bf r} \cdot \hat{\cal E}({\bf r}),
\label{H} \end{equation}
where $H_{ph}$ and $H_{el}$ denote the free-photon and free-electron
Hamiltonians, respectively, and $H_I$
is the photon-electron interaction in the dipole approximation.
Since  ${\bf u}({\bf r})$ varies on the scale of the photon wavelength,
$\hat{\cal E}({\bf r})$ does not vary appreciably on the scale of
the electron Fermi wavelength.
As a result, $H_I$ induces only an ``adiabatic change''
in the state vector \cite{Spra},
and we can show that the final state is simply given by \cite{Spra}
\begin{equation}
|\Psi' \r = \sum_n a_n e^{i \theta_n} |n \r |\varphi_a \r.
\label{SQWfinal} \end{equation}
where
\begin{equation}
\theta_n = \zeta n + \mbox{terms independent of $n$}.
\label{theta} \end{equation}
Here, $\zeta$ is an effective coupling constant which is a function of
the structural parameters such as the effective mass $m^*$
and the wire width:
\begin{equation}
\zeta =
{
2 \pi \hom  |\bra \vpb | ez | \vpa \ket|^2/\Delta
\over
\hbar^2 k / m^*
}
\int_{-\infty}^{\infty}
|u(x, y_0, z_0)|^2
dx,
\label{zeta} \end{equation}
where $y_0, z_0$ denote the center position of the wire
(which extends along the x axis), and
\begin{equation}
\Delta \equiv \epsilon_b - \epsilon_a - \hbar \omega
\label{Delta} \end{equation}
is the detuning energy. ($\epsilon_a$ and $\epsilon_b$ are the first and
the second subband energies.)

We see that {\it the final state
acquires $n$-dependent phase shift, $\theta_n$}.
If we could measure $\theta_n$ we would be able to know the photon number $n$.
However,
for the {\it single} quantum-wire structure as we are assuming in this section,
there is no way to measure $\theta_n$.
In the case of $a_n = \delta_{n,n_0}$, for example, $\theta_{n_0}$
is the {\it absolute} phase of the wavefunction, which is not a physical
quantity, and thus is unable to observe.
We therefore see that we could {\it not} measure $n$ if we used
a {\it single}-wire structure.
\bigskip\medskip

\noindent{\bf 5. Double quantum-wire structure}
\bigskip

\noindent
We now turn to the case of Fig.1;
a double-wire structure composed of
narrow (N) and wide (W) quantum wires.
As before, suppose that an electron wavepacket is emitted from the source.
As it proceeds towards the positive x direction,
the electron wave is split into two, and
the state vector of the coupled photon-electron system becomes
\begin{equation}
|\Psi \r = |\psi_{ph} \r |\psi_{el}\r = \sum_n a_n  |n \r
(|\varphi_a^N \r +  |\varphi_a^W \r)/\sqrt{2},
\label{initial} \end{equation}
where $\varphi_a^N(z)$ and $\varphi_a^W(z)$ denote
the lowest-subband eigenfunctions of the N and W wires, respectively.

Similarly to Eq.\ (\ref{SQWfinal}), the final state is shown to be \cite{Spra}
\begin{equation}
|\Psi' \r =
\sum_n a_n  |n \r
(e^{i \theta_n^N} |\varphi_a^N \r + e^{i \theta_n^W}  |\varphi_a^W \r)/\sqrt{2}.
\label{final} \end{equation}
where $\theta_n^{N,W}  = \zeta_{N,W} \, n \ +$ terms independent of $n$.
Here, $\zeta_N$ and $\zeta_W$ are the effective coupling constants of
the N and W wires, respectively, which are given by
Eq.\ (\ref{zeta}) with $\phi_{a,b} \rightarrow \phi_{a,b}^{N,W}$,
$\Delta \rightarrow \Delta_{N,W}$,  and
$y_0, z_0 \rightarrow y_0^{N,W}, z_0^{N,W}$.

Unlike the {\it absolute} phase in Eq.\ (\ref{SQWfinal}),
we can measure the {\it relative}  phase $\theta_n^N - \theta_n^W$ in
Eq.\ (\ref{final}) by the method  described in  the next section.
The relative phase is given by
\begin{equation}
\theta_n^N - \theta_n^W =  g n
\end{equation}
where $g \equiv \zeta_N - \zeta_W$ is an overall effective coupling constant.
Since the intersubband transition energy is higher in the N wire than in the W wire,
the detuning energies have opposite signs: $\Delta_N>0$, $\Delta_W < 0$,
as seen from Eq.\ (\ref{Delta}).
This results in $\zeta_N>0$, $\zeta_W < 0$, hence
$g = |\zeta_N|+|\zeta_W| \neq 0$.
(Typically, $\zeta_W \simeq - \zeta_N$, so that $g \simeq 2 \zeta_N$.)
Measurement of the relative phase thus provides us with the knowledge
about $n$ (see Eq.\ (\ref{J}) below).

Since $\zeta_N \neq \zeta_W$ (i.e., $g \neq 0$) is essential to the above
discussion, we also see that
a {\it double}-wire structure composed of  {\it identical} quantum wires
would not work as a photodetector.
Hence, the use of {\it double}-wire structure composed of
{\it non-identical} wires is essential to
the present QND photodetector.
\bigskip\medskip

\noindent{\bf 6. Measurement of the relative phase}
\bigskip

\noindent
We can measure the relative phase $\theta_n^N - \theta_n^W$ in
Eq.\ (\ref{final}) by composing an electron interferometer.
In Fig.1, a simple interferometer is employed:
The two phase-shifted components of Eq.\ (\ref{final}) is superposed
at the ``mode converter'', which is composed of a thin barrier of
50 \% transmittance.
The mode converter plays the same role as the beam splitter does
for the optical beam:
the input electron waves are superposed,
so that the state vector evolves into
\begin{equation}
|\Psi'' \r =
\sum_n a_n  |n \r
(C_{n+} |\varphi_+ \r + C_{n-}  |\varphi_- \r),
\label{final2} \end{equation}
where $|\varphi_\pm \r$ are the traveling modes of the two
output channels, and
\begin{equation}
C_{n \pm} =
[ e^{i (\theta_n^N + \theta_0)} \pm e^{i (\theta_n^W - \theta_0)} ]/2.
\label{C} \end{equation}
Here,
the additional phase angle $\theta_0$ is a function of
the structural parameters, such as the height and thickness of the barrier,
of the mode converter.

We measure the intensities of the output electron waves as the interference currents,
$J_+$ and $J_-$.
Equations (\ref{final2}) and (\ref{C}) yield
\begin{equation}
\l J_\pm \r
\propto \sum_n |a_n|^2 |C_{n \pm}|^2
= {1 \over 2} \sum_n |a_n|^2 [  1 \pm \cos (g n + \theta_0) ]
= {1 \over 2} [1 \pm \l \cos (g n + \theta_0)  \r].
\label{J} \end{equation}
When the mode converter is designed in such a way that $\theta_0 = - \pi /2$,
for example,  this relation yields
$\l J_+ \r - \l J_- \r \propto \l \sin g n \r$.
We can therefore measure $n$ by measuring $J_\pm$.
\bigskip\medskip

\noindent{\bf 7. QND property}
\bigskip

\noindent
As we will see in section 9,
we need {\it many} electrons to reduce the measurement error.
The many-electron versions of Eqs.\ (\ref{initial}), (\ref{final})
and (\ref{final2}) are obtained by taking
their Slater determinant for the electron part.
(Here, each electron state must of course be different in either of spin,
or the center position of the wavepacket, etc.)
Since we measure $J_\pm$ of such a many-electron state,
the state vector after the measurement is ``reduced'' to
an eigenstate of the many-electron $J_\pm$. (See also section 9.)
For the reduced state vector, only the photon part is of our interest.
When $N_\pm$ electrons are found in
the $\pm$ channels,  {\it the photon state after the measurement} is found to be
\begin{equation}
| \psi''_{ph}(N_+, N_-) \ket
=
    \left[ P(N_+, N_-)/{N \choose N_+} \right]^{-1/2}
    \sum_n a_n (C_{n+})^{N_+}(C_{n-})^{N_-} | n \ket,
\label{phfinal} \end{equation}
where $P(N_+, N_-)$ is the probability of finding  $N_\pm$ electrons
in the $\pm$ channels (for a given $N = N_+ + N_-$),
and is evaluated to be
\begin{equation}
P(N_+, N_-)
 = {N \choose N_+} \sum_n |a_n|^2 |C_{n+}|^{2 N_+}|C_{n-}|^{2 N_-}.
\label{prob} \end{equation}
In other words, with the probability of $P(N_+, N_-)$ the post-measurement
photon state becomes $| \psi''_{ph}(N_+, N_-) \ket$.
Let us confirm the QND property using these equations.

We first consider the case where
the initial photon state is a number state;
$| \psi_{ph} \r = | n_0 \r$.
Since $a_n = \delta_{n, n_0}$ in this case, we find from Eqs.\ (\ref{phfinal}) and
(\ref{prob}) that
$| \psi''_{ph}(N_+, N_-) \ket = | n_0 \r$.
That is,
when the pre-measurement photon state is a number state,
the post-measurement state becomes the {\it same} number state ---
no change occurs  by the measurement
either in the photon number or in the state vector!
Hence the name a QND photodetector \cite{qndquadra}.

We next consider the general case where
the initial photon state is given by Eq.\ (\ref{phinit}) with $a_n$ being
arbitrary.
In this case,
the {\it general}
requirement of quantum mechanics {\it requires}
some changes in the state vector. Otherwise, the uncertainty principle, for
example, would be broken. (See the next section.)
Therefore, even when you use a QND detector the state vector of the
measured system must be changed \cite{qndquadra,Spra,SF}.
Indeed, the post-measurement photon state, Eq.\ (\ref{phfinal}),
is clearly different from the initial state.
In particular,
the photon-number distribution after the measurement is
\begin{equation}
|\l n |  \psi''_{ph}(N_+, N_-) \r |^2 =
{
|a_n|^2 |C_{n+}|^{2 N_+}|C_{n-}|^{2 N_-}
\over
\sum_m |a_m|^2 |C_{m+}|^{2 N_+}|C_{m-}|^{2 N_-}
},
\label{dsteach} \end{equation}
which is {\it different} from that before the measurement,
$|\l n |  \psi_{ph} \r |^2 = |a_n|^2$.

However, the unique property of a QND detector can be seen
by considering an {\it ensemble} of many
equivalent systems \cite{qndquadra,Spra,SF}
--- such an ensemble has been very frequently used
(either explicitly or implicitly) in discussions on quantum physics \cite{vN}.
{\it For each member} in the ensemble, the above equations can be applied.
We can therefore calculate the density operator
$\hat \rho$ of the ensemble as follows.
Here, I will present the
density operator traced over the electron degrees of
freedom, $\hat \rho_{ph} = {\rm Tr}_{el} [\hat \rho]$, which
is of our principal interest.
Before the measurement, all members have the same state vector of
Eq.\ (\ref{phinit}), hence
\begin{equation}
\hat \rho_{ph} = \sum_{m,n} a_m a_n^* |m \r \l n |,
\label{rhoinitial} \end{equation}
and the photon-number distribution {\it over the ensemble}, ${\rm Prob}(n)$,
is simply given by ${\rm Prob}(n) = |a_n|^2$.
After the measurement, on the other hand,
a member in the state of Eq.\ (\ref{phfinal}) is found in the ensemble with
the probability of Eq.\ (\ref{prob}).
Therefore, the photon density operator (for  a given $N = N_+ + N_-$)  becomes
\begin{eqnarray}
\hat \rho''_{ph}
&=&
\sum_{N_+} P(N_+, N_-) | \psi''_{ph}(N_+, N_-) \r \l \psi''_{ph}(N_+, N_-) |  \nonumber \\
&=&
\sum_{n, m}
    a_m a_n^* | m \ket \bra n |
    \left( \hf e^{i \zeta_N (m - n)}
         + \hf e^{i \zeta_W (m - n)} \right)^N
\label{rhofinal} \end{eqnarray}
and the distribution after the measurement is
\begin{eqnarray}
{\rm Prob}''(n)
&=& \sum_{N_+} P(N_+, N_-) |\l n |  \psi''_{ph}(N_+, N_-) \r |^2  \nonumber \\
&=& \sum_{N_+} {N \choose N_+} |a_n|^2 |C_{n+}|^{2 N_+}|C_{n-}|^{2 N_-} \nonumber \\
&=&  |a_n|^2 (|C_{n+}|^2 + |C_{n-}|^2) = |a_n|^2,
\label{dstfinal} \end{eqnarray}
where use has been made of Eqs.\ (\ref{C}), (\ref{prob}) and (\ref{dsteach}).
We thus find that
{\it the photon-number distribution over the ensemble is unchanged}.
In this sense the QND photodetector is said to
cause no change in the ``statistical
distribution'' of the photon number, or,
to cause no ``backaction'' on the photon number \cite{qndquadra,Spra,SF}.
In particular, we find from Eq.\ (\ref{dstfinal}) that
the final state has the same average and variance of $n$
as the initial state:
\begin{equation}
\l n \r_{final} = \l n \r_{init},
\quad
\l \dns \r_{final} = \l \dns \r_{init }.
\label{nfinal} \end{equation}
This is in a sharp contrast with
conventional photodetectors, which drastically alter the photon-number
distribution by absorbing photons.

Note that all the above results referred to either the initial or the
final state.
It can be shown that {\it ${\rm Prob}(n)$
does change during the measurement}, i.e.,
during the photon-electron collision \cite{Spra,SF}.
The absence of change is
claimed {\it only for the post-measurement state}, and
{\it this suffices  to claim the QND property} \cite{Spra,SF}.
This is in a sharp contrast with previous QND photodetectors
\cite{qndquadra}, which were
claimed to cause no change of ${\rm Prob}(n)$ {\it throughout the measurement}.
This fact demonstrates that the operation principle of
the present QND photodetector is much different from the previous ones.
We recently developed a general theory which clarifies
the physics of various types of QND measurement \cite{SF}.
\bigskip\medskip

\noindent{\bf 8. Backaction noise generated by the measurement}
\bigskip

\noindent
We have seen that our QND photodetector causes no backaction on
the {\it measured variable} --- the photon number $n$,
in the sense of Eqs.\ (\ref{dstfinal}) and (\ref{nfinal}).
On the other hand, we expect from the uncertainty principle that
the detector {\it must} cause some backaction on
the phase $\phi$ --- the conjugate variable of $n$ --- of the photon
field \cite{qndquadra,Spra}.

To demonstrate this, we consider the case where the initial photon state
is a coherent state $| \xi \r$, for which
\begin{equation}
a_n =  e^{- |\xi|^2/2} { {\xi^n}/{\sqrt{n !}} },
\label{coherent} \end{equation}
which yields $\l n \r_{init} = \l \dns \r_{init} = |\xi|^2$, and
the phase fluctuations are evaluated to be
$\l \delta \phi^2 \r_{init} \simeq 1/4 |\xi|^2$ for large $|\xi|$ \cite{Loudon}.
(In the large-$|\xi|$ limit, in particular,
both $\l \dns \r_{init}/(\l n \r_{init})^2$ and
$\l \delta \phi^2 \r_{init}$ tend to zero,
and $| \xi \r$ approaches the classical state in which
$\hat a$ in Eq.\ (\ref{E}) is replaced with $\xi$ \cite{Loudon}.)
It is convenient to introduce  ``quadrature variables,''
$\hat a_1$ and $\hat a_2$, which correspond to the amplitudes of
the cosine and sine parts of the optical field \cite{Loudon};
\begin{equation}
\hat a_1 \equiv (\hat a + \hat a^\dagger)/2,
\quad
\hat a_2 \equiv (\hat a - \hat a^\dagger)/2i.
\end{equation}
The above fluctuations in $n$ and $\phi$ are translated into
fluctuations of these variables as
$\l \delta a_1^2 \r_{init} = \l \delta a_2^2 \r_{init} = 1/4$.
Therefore, in the $a_1$-$a_2$ plane
a coherent state can be represented as a circular
``cloud'' \cite{Loudon}, which schematically visualizes the fluctuations,
as shown in Fig.~2 (a).
In this diagram, $n$ corresponds to the square of the radial
distance from the origin,
and $\phi$ to the azimuthal angle \cite{Loudon}.
Fluctuations in $n$ and $\phi$ are therefore represented by
the spread of the cloud in the radial and azimuthal directions, respectively.

\begin{figure}[p]
\vspace{145mm}
\noindent
{\bf Fig.~2}
(a) When the initial photon state is a coherent state, it can be represented as a
circular ``cloud'' in the $a_1$-$a_2$ plane.
(b) When two quantum wires are identical,
the photon state rotates by $\zeta$ after the collision with an electron
in the wires.
(c)-(f)  When two quantum wires are non-identical, on the other hand,
the photon state is drastically deformed.
(c), (d), (e) and (f) represent the photon state after
the collision with one, two, three and many electrons, respectively.
In (b)-(e) a large value of $\zeta$ is assumed
in order to  make the diagrams vivid, whereas
realistic small $\zeta$ is assumed in (f).
\end{figure}

It is instructive to consider first the case of
{\it identical wires}, for which $\zeta_N = \zeta_W$ ($\equiv \zeta$).
In this case Eqs.\ (\ref{rhofinal}) and (\ref{coherent}) yield
\begin{equation}
\hat \rho''_{ph} =
\sum_{n, m} a_m a_n^* e^{i N \zeta (m-n)}| m \ket \bra n |
= | e^{i N \zeta} \xi \r \l e^{i N \zeta} \xi |,
\end{equation}
where, as before, $N$ denotes the number of colliding electrons.
Thus, the identical wires just induce
the phase rotation in the parameter $\xi$, and the
final photon state is the same coherent state as the initial state except for this
unimportant phase rotation.
This is illustrated in Fig.~2 (b) for the case of $N=1$.

For non-identical wires, on the other hand,
$\zeta_N \neq \zeta_W$, and
$\hat \rho''_{ph}$ can no longer be factorized in such a simple form.
In particular, off-diagonal terms,
$\l m | \hat \rho''_{ph} | n \r$ with $m \neq n$,
are significantly reduced with increasing $N$.
This leads to phase randomization because the
quantum-mechanical phase is a measure
of the off-diagonal coherence.
In fact, we can show for large $|\xi|$ that \cite{Spra}
\begin{equation}
\l \delta \phi^2 \r_{final}
= \l \delta \phi^2 \r_{init} + \ban^2,
\quad \ban^2 \simeq {{N g^2}/ 4},
\label{ban} \end{equation}
where, as before, $g \equiv \zeta_N - \zeta_W$.
The physical origin of this backaction noise, $\ban^2$, is
sketched in Fig.~2 (c)-(f), where for simplicity
$\zeta_N = - \zeta_W$ ($\equiv \zeta$) is assumed.
When one electron collides with the photons,
the electron amplitudes in the two wires simultaneously cause rotations
of angles $\zeta_N = \zeta$ and $\zeta_W = -\zeta$, as shown in Fig.~2 (c).
As a result, the photon state is split into two clouds.
When one more electron joins the game,
each cloud is again split into two, and
the photon state becomes as Fig.~2 (d).
Similarly, we get Fig.~2 (e) for $N=3$, and finally Fig.~2 (f) for large $N$.
This banana-like state is a graphical representation of
$\hat \rho''_{ph}$, Eq.\ (\ref{rhofinal}).
As compared with the initial state (a),
we see that the final state (f) indeed has larger
phase fluctuations (which correspond to the
azimuthal distribution), while the magnitude of the
photon-number fluctuations
(the radial distribution) remains the same.
Comparison between (b) and (f)  demonstrates that
the pair of non-identical wires,
for which $\zeta_N  \neq \zeta_W$, is the very origin of the
backaction noise, $\ban^2$.
\bigskip\medskip

\noindent{\bf 9. Measurement error}
\bigskip

\noindent
A principal postulate of quantum mechanics
is that when  ``ideal measurement'' is performed
the state vector of the measured
system is reduced to an eigenstate of the measured variable.
For photon-number measurement, for example,
ideal measurement would lead to the post-measurement
density operator of the form,
\begin{equation}
\hat \rho^{ideal \ meas}_{ph} =
\sum_{n} |a_n|^2  | n \ket \bra n |.
\label{ideal} \end{equation}
Actual measuring devices, however, are non-ideal
in two points: (i) they would destroy (demolish)
the photon state by, say, absorbing photons, and
(ii) they have a finite measurement error.
Therefore, the density operator (or the state vector)
will be reduced to another form.
Theory and experiment of such non-ideal measurement have been attracting
much attention recently \cite{G,SF,UK}.

The present QND photodetector is a good example to understand
the physics of non-ideal measurement.
Although the QND photodetector does not absorb photons,
it may be non-ideal because of a finite measurement error, and
the state vector would not be reduced completely.
Let us examine this subject, as well as the origin
of the measurement error.

Measurement consists of a series of physical interactions
which occur among many degrees of freedom in
the measured system and the measuring apparatus \cite{vN}.
In our case, the interactions consist of
that between photons and electrons,
that between the electrons and ammeters which measure $J_\pm$,
that between the ammeters and a recorder which records the values of $J_\pm$,
and so on.
The key to treat non-ideal measurement is
the fact that among these interactions
we can (almost always) find an interaction process
which can be approximated as ideal measurement---
for such a process we can apply the above principal postulate
of quantum mechanics, and everything can then be evaluated
(at least in principle) \cite{SF}.
Note that we do not need any additional postulate;
{\it we can treat non-ideal measurement within the
standard framework of quantum mechanics} \cite{SF}.

In our case, we have assumed in section 7 that
the measurement of the {\it electronic current}, $J_\pm$, is ideal.
As a result, the state vector of the coupled
photon-electron system is reduced to an eigenstate of $J_\pm$.
The photon part of the reduced state vector is shown in
Eq.\ (\ref{phfinal}), and the reduced density operator in Eq.\ (\ref{rhofinal}).
The photon number $n$ is {\it estimated} from the measured values of $J_\pm$
through Eq.\ (\ref{J}).
Since we get information on $n$, the whole process can be called
measurement of $n$.
That is, we measure $n$ through ideal measurement of $J_\pm$.
This measurement of $n$ is non-ideal because
it has a finite measurement error.
In fact,
since $J_\pm$ are quantum interference currents,
they have finite quantum fluctuations
\cite{mbint}--\cite{SS}, which make
the estimation of $n$ ambiguous \cite{Spra,SS,remark}.
(This is a quite general result for quantum interference
devices, as shown in \cite{SS,remark}.)
Namely, the fluctuations of $J_\pm$ give rise to
a finite error in measurement of $n$,
and the present QND device works as
a  {\it non-ideal} measuring device of $n$.
As a result, the post-measurement photon state is
{\it not completely reduced} to an eigenstate of $n$, as
explicitly seen from
Eq.\ (\ref{rhofinal}),  which shows that
$\hat \rho''_{ph} \neq \hat \rho^{ideal \ meas}_{ph}$ for finite $N$.

As $N$ is increased
$\hat \rho''_{ph}$ approaches $\hat \rho^{ideal \ meas}_{ph}$.
We thus expect that
the measurement error  decreases with increasing $N$.
This is indeed the case; the measurement error
is evaluated to be \cite{Spra}
\begin{equation}
\me = {1 /{g^2 N} }.
\label{me} \end{equation}
This result can be understood as follows.
As the effective coupling $g$ is increased,
the flow of information from the light field to
the electrons is increased, hence $\me \propto 1/g^2$.
On the other hand, we get to know the photon number by
measuring the electron phase shift.
To measure the phase shift, however, we
need many electrons  because of
the number-phase uncertainty principle (of electron waves) \cite{mbint,SS}.
This results in $\me \propto 1/N$.
It was shown in Refs.\ \cite{SS} that similar discussions
can be applied to most
quantum interference devices, and their
fundamental limits have been derived \cite{remark}.

Interestingly, if we multiply $\me$ by
the backaction noise
$\ban^2$, Eq.\ (\ref{ban}),  we get a constant;
$\me \ban^2 \simeq 1/4$, whereas the number-phase uncertainty principle
(of a light field)
gives $\me \ban^2 \geq 1/4$ \cite{Loudon}.
This means that the present device is a very effective
measuring device in the sense that it extracts the information
on the measured variable $n$
with the minimum cost of the backaction noise in the
conjugate variable $\phi$.
\bigskip\medskip

\noindent{\bf 10. Summary}
\bigskip

\noindent
I have analyzed
a quantum non-demolition (QND) photodetector composed of
a double quantum-wire electron interferometer, which
measures the photon number $n$ without absorbing photons
(more precisely,
without changing distribution of $n$).
It is shown that
(i) If we used an {\it single}-wire structure, or if
we used a double-wire structure composed of two {\it identical}  wires,
we could not get information on the photon number.
It is therefore essential to use a double-wire structure composed of
{\it non-identical} wires.
(ii) Such a double-wire structure, on the other hand,
is the very origin of the backaction noise, which
appears as an increase of quantum
fluctuations of the {\it phase} of the light field.
(iii) The QND photodetector works as a {\it non-ideal} measuring device
because it has a finite measurement error.
As a result, the photon state is not completely reduced
to an eigenstate of $n$.
(iv) The measurement error, $\me$,
comes from quantum fluctuations of electrical currents in the interferometer.
Because of this fluctuation, we need many electrons to measure
the phase shift which is
induced by the light field. As a result, $\me \propto 1/N$,
where $N$ denotes the number of colliding electrons.
(v) The error is also inversely proportional
to an effective coupling constant $g^2$
between photons and electrons. The coupling constant is a function of the
structural parameters of the quantum wires.
(vi) The measurement error and the backaction noise is
closely related: $\me \ban^2 \simeq 1/4$.
Namely, the backaction noise is proportional to $N g^2$ and is also
a function (through $g^2$) of the structural parameters of the wires.

These results demonstrate close relationships between measurement
and fluctuations, and not only shed light on the physics of quantum
measurement, but also suggest fundamental limitations and possibilities
of nanostructure devices.


\begin{thebibliography}{99}
\bibitem{unitary}
More precisely, the relation between the old and new $\hat a$'s are
expressed as a unitary transformation, and the commutation relations are
preserved.
\vspace{-2mm}
\end{thebibliography}
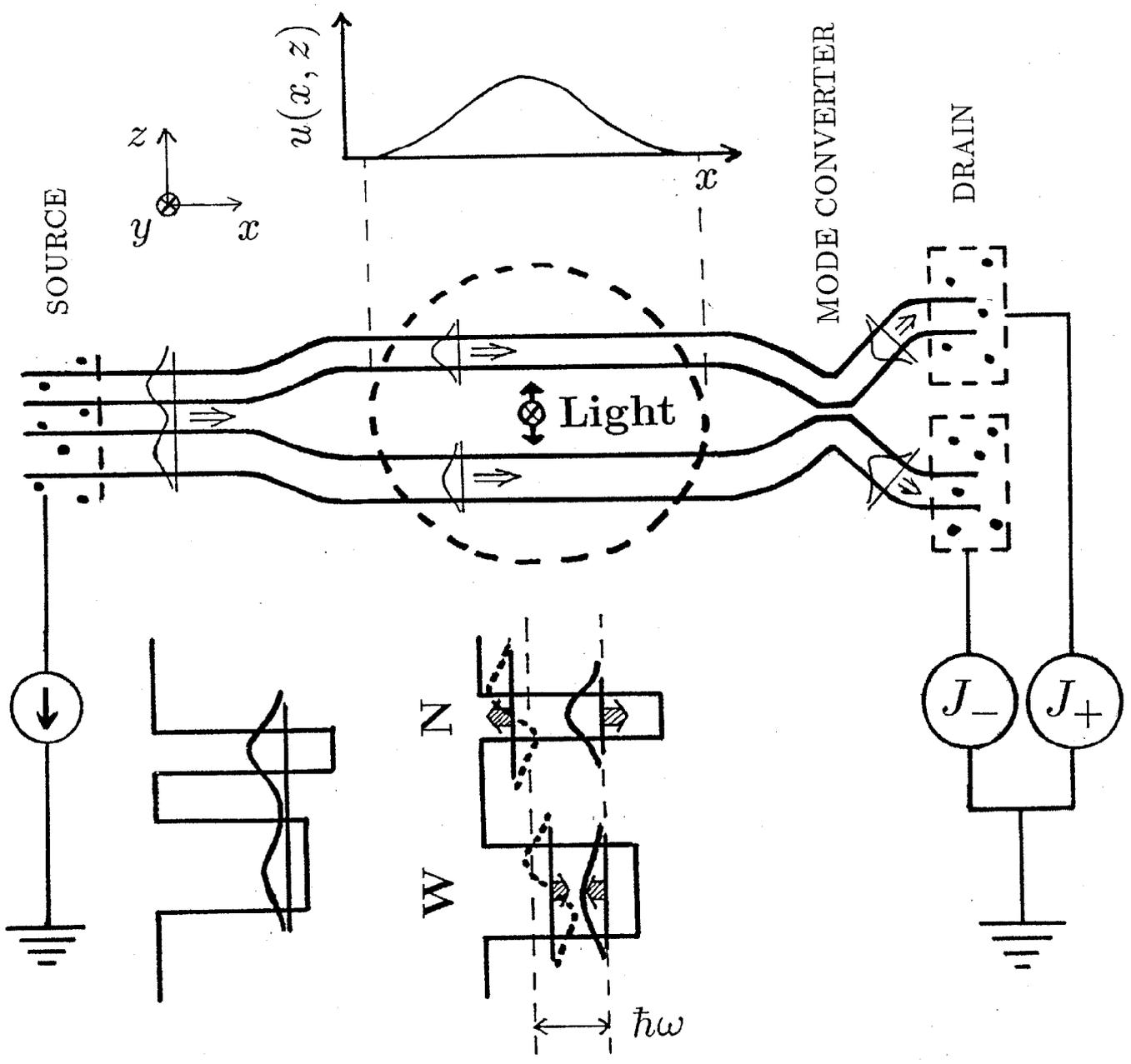
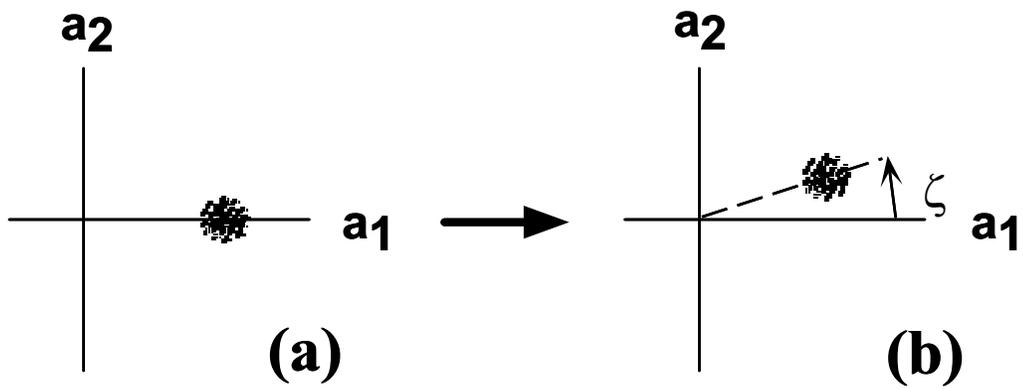
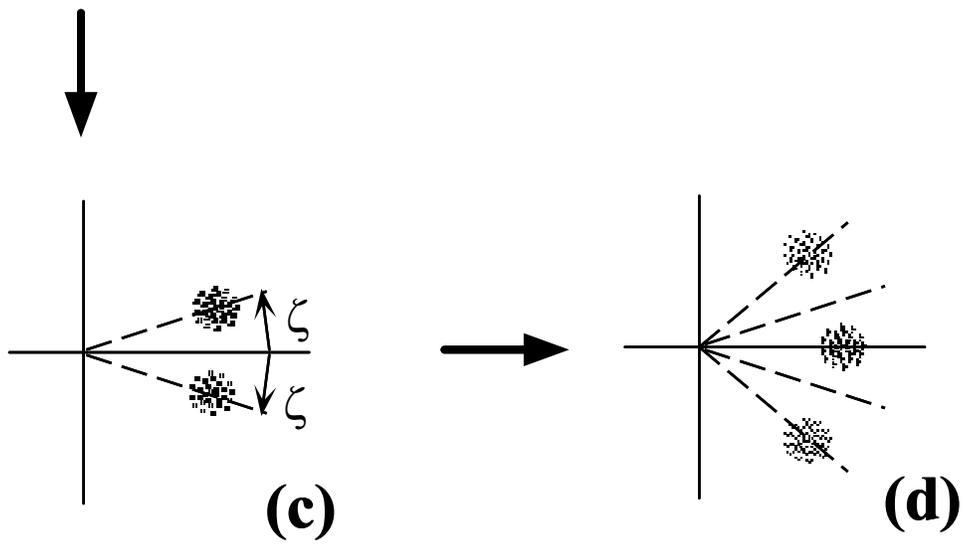
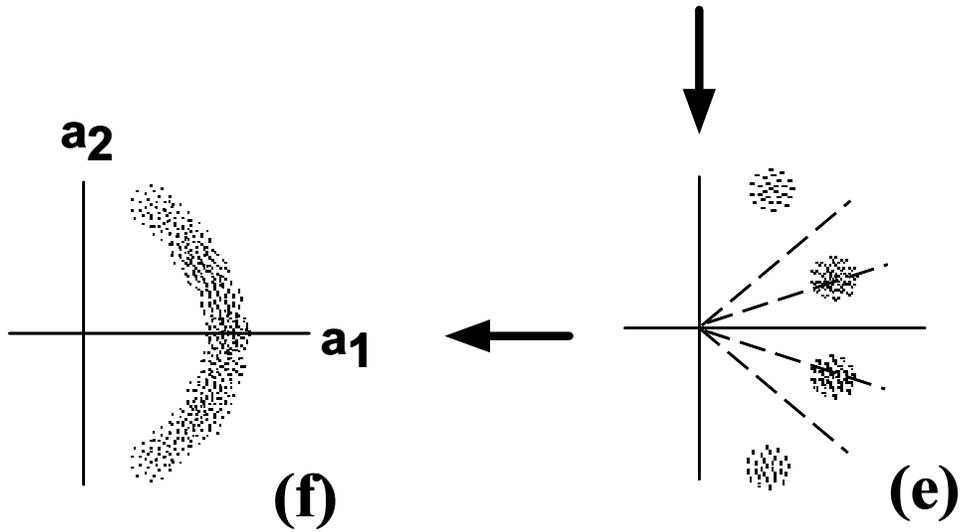
\end{document}